\documentstyle[11pt,newpasp,twoside,epsf]{article}
\begin{document}

\title{Monte Carlo Simulations of Globular Cluster Dynamics}
 
\author{Frederic A.\ Rasio}
\affil{Dept of Physics, MIT, Cambridge, MA 02139, USA}

\begin{abstract} We have developed a new parallel supercomputer code
based on H\'enon's Monte Carlo
method for simulating the dynamical evolution of globular clusters. 
This new code allows us to calculate
the evolution of a cluster containing a realistic number of stars
($N\sim 10^5-10^6$)  in about a day of computing time. The discrete, star-by-star
representation of the cluster in the simulation allows us to treat
naturally a number of important processes, including
single and binary star evolution, all dynamical interactions of single
stars and primordial binaries, and tidal interactions with the Galaxy.
\end{abstract}

\section{Introduction}

The first Monte Carlo methods for calculating the dynamical evolution 
of star clusters
in the Fokker-Planck approximation were developed more than 30 years ago. They
were first used to study the development of the gravothermal instability 
(H\'enon 1971a,b; Spitzer \& Hart 1971a,b). 
More recent implementations have established the Monte Carlo method
as an important
alternative to direct $N$-body integrations (see Spitzer 1987 for an
overview, and the article by Giersz in this volume).
The main motivation for our recent work at MIT was our realization a few
years ago that the  latest generation of
parallel supercomputers now make it possible
to perform Monte Carlo simulations for a number of objects equal 
to the actual number of stars in a globular cluster (in contrast,
earlier work was limited to using  a small number of representative
``superstars,'' and was often plagued by high levels of numerical noise).
Therefore, the Monte Carlo method allows us to do right now what remains
an elusive goal for $N$-body simulations (see, e.g.,  Aarseth 1999, and
the article by Makino in this volume): perform realistic, star-by-star
computer simulations of globular cluster evolution.
Using the correct number of stars in a dynamical simulation
ensures that the relative rates of different dynamical processes (which all
scale differently with the number of stars) are correct. This is
particularly crucial if many different dynamical processes are to be 
incorporated, as must be done in realistic simulations
(cf.\ the article by Heggie in this volume).

Our implementation of the Monte Carlo method is described in detail in
the papers by 
Joshi, Rasio, \& Portegies Zwart (2000), Joshi, Nave, \& Rasio (2000), and
Joshi \& Rasio (2000).
We adopt the usual assumptions of spherical symmetry (with a 2D phase space 
distribution function $f(E,J)$, i.e., we do {\it not\/} assume isotropy)
and standard two-body relaxation in the weak scattering limit (Fokker-Planck 
approximation). In its simplest version, our code computes the dynamical 
evolution of a self-gravitating spherical cluster of $N$ point masses 
whose orbits 
in the cluster are specified by an energy $E$ and angular momentum $J$, 
with perturbations $\Delta E$ and $\Delta J$ evaluated on a timestep that 
is a fraction of the local two-body relaxation time. The cluster is assumed to
remain always very close to dynamical equilibrium (i.e., the relaxation time
must remain much longer than the dynamical time). We have performed
a large number of test calculations and comparisons with direct $N$-body
integrations, as well as direct integrations of the Fokker-Planck equation in phase 
space, to establish the accuracy of our basic treatment of two-body relaxation
(Joshi et al.\ 2000a). Fig.~1 shows the results from a typical comparison 
between Monte Carlo and $N$-body simulations.
Our main improvements over H\'enon's original method
are the parallelization of the basic algorithm and the development of a more 
sophisticated method for determining the timesteps 
and for computing the two-body relaxation from representative encounters
between neighboring stars. Our new method allows the timesteps to be made 
much smaller in order to resolve the dynamics in the cluster core more 
accurately.

\begin{figure}
\plotfiddle{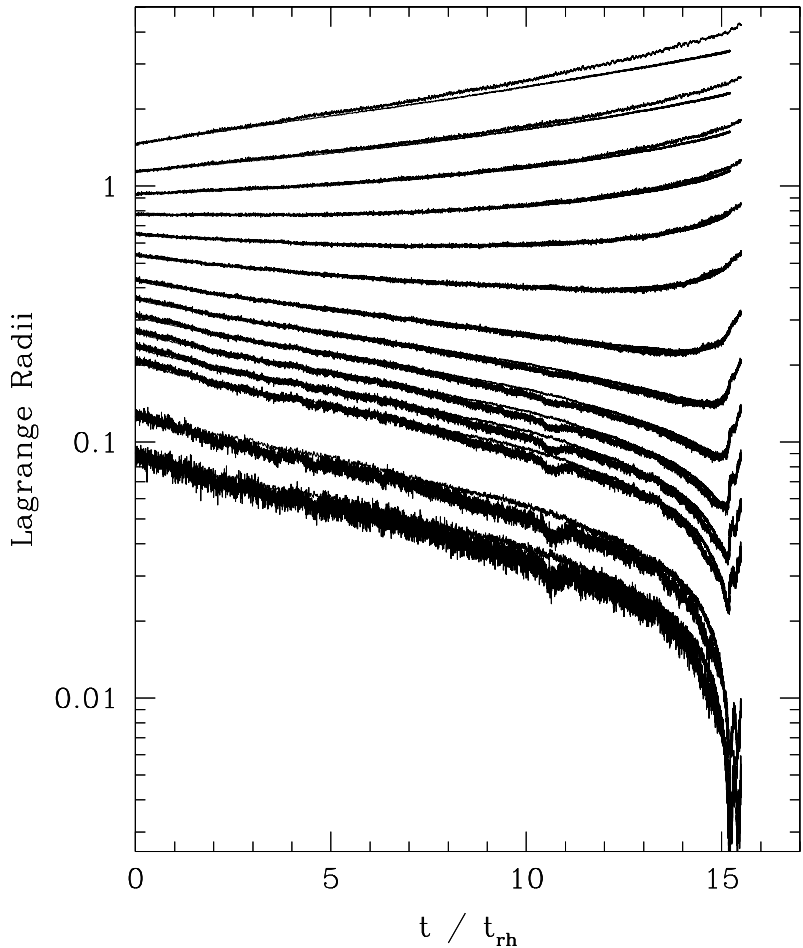}{13 cm}{0}{65}{65}{-210}{-100}
\caption{Evolution of the Lagrange radii for an isolated, single-component
Plummer model
(from bottom to top: radii containing 0.35\%, 1\%, 3.5\%, 5\%, 7\%, 10\%, 
14\%, 20\%, 30\%, 40\%, 50\%, 60\%, 70\%, and 80\% percent of the 
total mass are shown as a function of time, 
given in units of the initial half-mass relaxation time).
The results from
a direct $N$-body integration with $N=16,384$ (noisier lines)
and from a Monte Carlo integration with $N=10^5$ stars (smoother lines)
are compared. The Monte Carlo simulation was completed in less than a day
on a Cray/SGI Origin2000 parallel supercomputer, while the $N$-body
integration ran for over a month on a dedicated GRAPE-4 computer. 
The agreement between the $N$-body and Monte Carlo results
is excellent over the entire range of Lagrange radii and time.  
The small discrepancy in the outer Lagrange radii is caused mainly
by a different treatment of escaping stars in the two models. 
In the Monte Carlo model, escaping stars are removed from the 
simulation and therefore not included in the determination of the 
Lagrange radii, whereas in the $N$-body model escaping stars are 
not removed. Note also that the Monte Carlo simulation is terminated 
at core collapse, while the $N$-body simulation continues beyond 
core collapse.}
\end{figure}

\section{Summary of Recent Results}

Our recent work has focused on the addition of more realistic stellar and binary
processes to the basic Monte Carlo code, as well as a simple but accurate 
implementation of a static tidal boundary in the Galactic field (Joshi et al.\
2000b). As a first application, we have studied the dependence on initial
conditions of globular cluster lifetimes in the Galactic environment.
As in previous Fokker-Planck studies (Chernoff \& Weinberg 1990;
Takahashi \& Portegies Zwart 1998),
we include the effects of a power-law initial mass function (IMF), mass loss 
through a tidal boundary, and single star evolution, and we
consider initial King models with varying central concentrations.
We find that the disruption and core-collapse times of our models are 
significantly longer than those obtained with previous 1D (isotropic) 
Fokker-Planck calculations, but agree well with more recent results
from direct $N$-body simulations and 2D Fokker-Planck integrations
(see also the article by Takahashi in this volume).
In agreement with previous studies,
our results show that the direct mass loss due to stellar evolution 
causes most clusters with a low initial central concentration 
to disrupt quickly in the Galactic tidal field. The disruption is
particularly rapid for clusters with a relatively flat IMF. 
Only clusters born with high central concentrations 
or with very steep IMFs 
are likely to survive to the present and undergo core collapse. 

In another recent study, we have used our Monte Carlo code to examine the
development of the Spitzer ``mass stratification instability'' in simple
two-component clusters (Watters, Joshi, \& Rasio 2000).
We have performed a large number of
dynamical simulations for star clusters containing two stellar populations 
with individual masses 
$m_1$ and $m_2 > m_1$, and total masses $M_1$ and $M_2 < M_1$.  We use both 
King and Plummer model initial conditions and we perform simulations for a wide
range of individual and total mass ratios, $m_2/m_1$ and $M_2/M_1$, in order to
determine the precise location of the stability boundary in this 2D parameter
space. As predicted originally by Spitzer (1969) using simple analytic arguments, we
find that unstable systems never reach energy equipartition, and are
driven to rapid core collapse by the heavier component.
These results have important implications for
the dynamical evolution of any population of primordial
black holes or neutron stars in globular clusters. In particular,
primordial black holes with $m_2/m_1\sim10$ are expected to undergo 
very rapid core collapse independent of the background cluster, and 
to be ejected from the cluster through dynamical interactions 
between single and binary black holes (see Portegies Zwart \& McMillan 2000
and references therein).
We have also used Monte Carlo simulations of simple two-component systems
to study the evaporation (or retention) of {\it low-mass\/} objects in globular clusters,
motivated by the surprising recent observations of planets and brown dwarfs in 
several clusters (Fregeau et al.\ 2000).

\begin{figure}
\plotfiddle{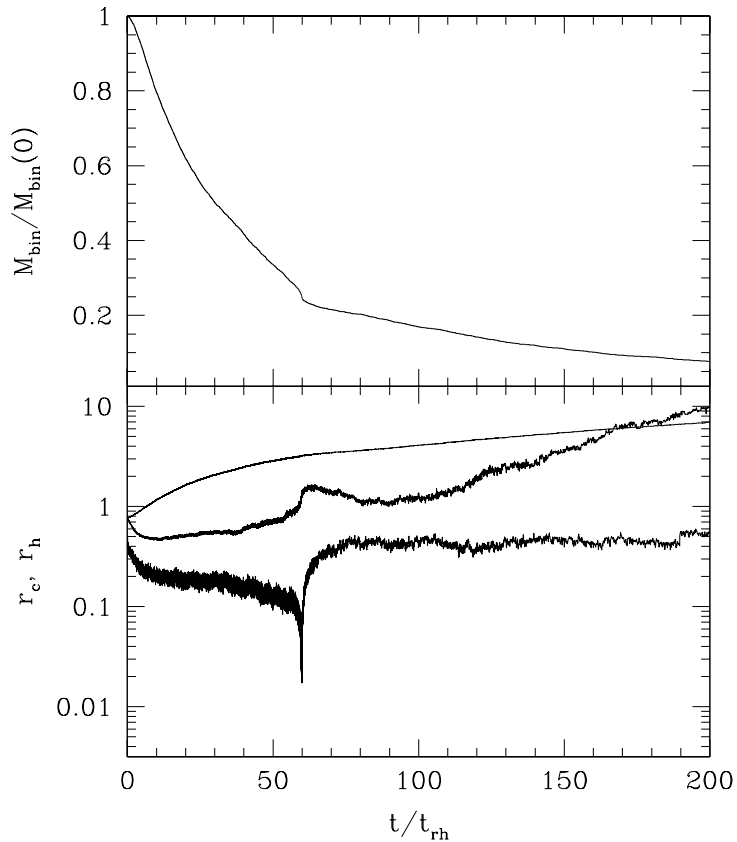}{11 cm}{0}{57}{57}{-180}{-80}
\caption{Results of a Monte Carlo simulation for the evolution of an isolated
Plummer model containing $N=3\times10^5$ equal-mass stars, with 10\% of the stars
in primordial binaries. The binaries are initially distributed uniformly throughout 
the cluster, and with a uniform distribution in the logarithm of the binding
energy (roughly between contact and the hard-soft boundary, i.e., no soft binaries
are included). The simulation
includes a treatment of energy production and binary destruction through
binary-single and binary-binary interactions. Stellar evolution and tidal
interactions with the Galaxy are not included.
Time is given in units of the initial half-mass relaxation time $t_{\rm rh}$.
The upper panel shows the evolution of the total mass (or number) of binaries.
The lower panel shows, from top to bottom initially, the half-mass radius
of the entire cluster, the half-mass radius of the binaries, and the cluster 
core radius. These quantities are in units of the virial radius of the cluster.
Note the long, quasi-equilibrium phase of ``binary burning'' 
lasting until $t\simeq60\,t_{\rm rh}$, followed by a brief
episode of core contraction and re-expansion to an even longer quasi-equilibrium
phase with an even larger core. By $t\sim 100\,t_{\rm rh}$, only about 
15\% of the initial population of binaries remains in the cluster, but this
is enough to support the cluster against core collapse for another
$\sim 100\,t_{\rm rh}$. For most globular clusters, $t_{\rm rh}\sim10^9\,$yr, and
this is well beyond a Hubble time. The evolution shown here should be contrasted
to that of an identical cluster, but containing single stars only (Fig.~1), where
core collapse is reached at $t\simeq15\,t_{\rm rh}$.}
\end{figure}

Much of our current work concerns the treatment of dynamical interactions with
primordial binaries. We are in the process of completing a first study of globular
cluster evolution with primordial binaries (Joshi \& Rasio 2000), based on the same 
set of approximate
cross sections and recipes for dynamical interactions
used in the Fokker-Planck simulations of Gao et al.\ (1991).
Typical results are illustrated in Fig.~2. The heating of the cluster core generated
by a small population of primordial binaries can support the cluster against core
collapse for very long times. 

The addition of binary stellar evolution processes will allow us to study in detail
the dynamical formation mechanisms for many exotic objects, such as X-ray binaries,
millisecond radio pulsars, and cataclysmic variables, which have been detected
in large numbers in globular clusters. For example, exchange interactions
between neutron stars and primordial binaries can lead to common-envelope
systems and the formation of short-period neutron-star / white-dwarf binaries
that can become visible both as ultracompact X-ray binaries and binary millisecond
pulsars with low-mass companions (see, e.g., Camilo et al.\ 2000, on observations
of 20 such millisecond radio pulsars in 47~Tuc; Rasio, Pfahl, \& Rappaport 2000 
present a preliminary study of this dynamical formation scenario, based on simplified
Monte Carlo simulations).

We are also currently working on
incorporating a more realistic treatment of tidal interactions, and, in particular,
tidal shocking through the Galactic disk (based on Gnedin, Lee, \& Ostriker 1999).
Tidal shocks can accelerate significantly both core collapse and the evaporation 
of globular clusters, reducing their lifetimes in the Galaxy (Gnedin \& Ostriker 1997).

Future work will include a fully dynamical treatment of all strong binary-single
and binary-binary interactions (exploiting the parallelism of the code to perform
separate numerical 3- or 4-body integrations for all dynamical interactions) as well
as a fully dynamical treatment of tidal shocking (performing short, $N$-body integrations
for each passage of the cluster through the Galactic disk or bulge).

\acknowledgements
This work is supported in part by NSF Grant AST-9618116, NASA ATP 
Grant NAG5-8460, and by an Alfred P.\ Sloan Research Fellowship.
Our computations are performed on the Cray/SGI Origin2000 
supercomputer at Boston University under NCSA Grant AST970022N.

\newpage

\subsubsection{Ivan King:}

You showed that binaries stabilize a cluster for much longer than a Hubble time.
Yet we see 15--20\% of clusters in a state of core collapse. How do you reconcile
this?

\subsubsection{Fred Rasio:}

Indeed, if the type of evolution shown in Fig.~2 applied to all globular
clusters, there would be no ``core-collapsed'' clusters in the Galaxy. 
However, the timescale
on which real clusters will exhaust their primordial binary supply and undergo
(deep) core collapse depends on a number of factors not considered here: the
initial primordial binary fraction (some clusters may have much fewer binaries
than the 10\% assumed in Fig.~2), the orbit of the cluster in the Galaxy
(the simulation of Fig.~2 is for an isolated cluster, but mass loss and
tidal shocking can accelerate the evolution dramatically), the stellar IMF
(the cluster shown in Fig.~2 contains all equal-mass stars and binaries;
a more realistic mass spectrum will also accelerate the evolution), etc.

However, the simple picture that emerges from Fig.~2 may well, to first approximation,
describe the dynamical state of most Galactic globular clusters observed today.
Note that, for a cluster in the stable ``binary burning'' phase of Fig.~2,
the ratio of half-mass radius to core radius $r_{\rm h}/r_{\rm c}\simeq
2-10$ (for $t\sim10\,$Gyr and $t_{\rm rh}\sim0.1-10\,$Gyr), which is precisely 
the range of values observed for the $\sim80\%$ of
clusters that have a well-resolved core and are well-fitted by King models.
Some of these clusters may have gone in the past through a brief episode of ``moderate core
collapse'' (as shown around $t\simeq60\,t_{\rm rh}$ in Fig.~2). Yet, I do not
believe that they should be called ``core-collapsed'' or ``post-core-collapse'' 
(nor would they be
classified as such by observers). Unfortunately some theorists will even call
``core-collapsed'' clusters that have just reached the {\it initial phase\/} of
binary burning ($t\simeq 10-50\,t_{\rm rh}$ in Fig.~2). Since the core has 
just barely contracted by a factor $\sim2-3$ by the time it reaches this
phase, it seems hardly justified to speak of a ``collapsed'' state.

\end{document}